\documentclass{article}

\usepackage[final]{neurips_2019}
\usepackage{dsfont}
\usepackage{pgfplots}
\pgfplotsset{width=6.25cm, compat=newest}
\usepgfplotslibrary{fillbetween}
\usepackage{siunitx}{
\sisetup{output-exponent-marker=\ensuremath{\mathrm{e}}}}

\usepackage[utf8]{inputenc} 
\usepackage[T1]{fontenc}   
\usepackage{hyperref}   
\usepackage{url}     
\usepackage{booktabs}   
\usepackage{amsfonts}   
\usepackage{nicefrac}   
\usepackage{microtype}  
\usepackage{graphicx}
\usepackage{blindtext}
\usepackage{wrapfig}
\usepackage{wrapfig,lipsum,booktabs}
\usepackage{subcaption}

\title{Faster Peace via Inclusivity: An Efficient Paradigm to Understand Populations in Conflict Zones}

\author{%
  Jordan Bilich \\
  Remesh\\
  \texttt{jordan@remesh.org} \\
   \And
   Michael Varga \\
   Remesh \\
   \texttt{michael@remesh.org} \\
   \AND
   Daanish Masood \\
   United Nations\\
   \texttt{masoodd@un.org} \\
   \And
   Andrew Konya \\
   Remesh \\
   \texttt{andrew@remesh.org} \\
}

\begin{document}

\definecolor{blue}{HTML}{1f77b4}
\definecolor{orange}{HTML}{ff7f0e}
\definecolor{green}{HTML}{2ca02c}

\maketitle

\vspace{-.65cm}
\begin{abstract}
United Nations practice shows that inclusivity is vital for mediation to be successful in helping end violent conflict and establish lasting peace. However, current methods for understanding the views and needs of populations during dynamic situations create tension between inclusivity and efficiency. This work introduces a novel paradigm to mitigate such tension. In partnership with collaborators at the United Nations we develop a realtime large-scale synchronous dialogue process (RLSDP) to understand stakeholder populations on an hour timescale. We demonstrate a machine learning model which enables each dialogue cycle to take place on a minute-timescale. We manage a  key risk related to machine learning result trustworthiness by computing result confidence from a fast and reliable estimation of posterior variance. Lastly, we highlight a constellation of risks stemming from this new paradigm and suggest policies to mitigate them.
\end{abstract}

\section{Introduction}

Violent conflict leads to deaths, poor health, degraded education, inequality, and economic loss. Since 2011, conflicts worldwide have killed up to 100,000 people a year [1], with indirect deaths estimated to be 3-15 times greater [2]. By 2030, over half of the world’s poor is projected to be living in countries affected by high levels of violence [1]. Migration, malnutrition, destroyed infrastructure, and distressed environments due to conflict lead to poor health [3], increased infant mortality [4], and decreases in the quality of childhood education [5]. Conflict tends to disproportionately impact those with lower socio-economic status, leading to increased economic inequality [6]. And, the overall economic losses due to conflict have doubled over the last decade to an estimated \$1.02 trillion per year (\$3 billion per day)  [7].

While ceasefires are helpful in preventing escalation once a conflict has begun, long-term peace often requires a successful dialogue and mediation process between the parties. Consequently, over \$27b is spent annually on peacebuilding efforts [7]. Article 33 of the Charter of the United Nations stresses that “the parties to any dispute, the continuance of which is likely to endanger the maintenance of international peace and security, shall, first of all, seek a solution by negotiation, enquiry, mediation, conciliation, arbitration, judicial settlement” [8]. UN practice shows that for a mediation and dialogue process to be successful, inclusivity is vital. The UN defines “inclusivity” as the extent and manner in which the views and needs of conflict parties and other stakeholders are represented and integrated into the process and outcome of a [conflict] mediation effort [9].  Given the large number of stakeholders in most conflicts, mediators are often hard-pressed for ways to poll the positions, needs, and interests of stakeholder populations. Further, the reality that the positions of these populations tend to shift, as the dynamics of a conflict or dialogue process evolve, presents a unique challenge. 

Consequently, mediators have to grapple with potential tensions between inclusivity and efficiency [10]. Additionally, while machine learning has the potential to increase efficiency, it is often viewed as too risky for high-stakes decisions  [11], suffering from a lack of result trustworthiness [12]. As a means to decreasing these tensions, in partnership with collaborators at the UN, we develop a realtime large-scale synchronous dialogue process (RLSDP) enabled by machine learning and manage risk related to result trustworthiness by computing result confidence from a reliable estimation of posterior variance. Further, we highlight risks stemming from this new paradigm and suggest policies to mitigate them.

\section{Methods}
\textbf{Dialgoue process.} Prior methods to gather the views and needs of stakeholder populations for inclusivity  include asynchronous dialogue platforms, surveys, focus groups, and social media analysis [10]. However, these methods manifest a tradeoff between conversational agility and statistical reliability. We aim to bridge this tradeoff by designing an approach with the agile dynamics of live conversation at a statistically relevant scale. 

We specify a RLSDP to be characterized by a continuous sequence of minute-scale exchanges between dialogue moderators and a large participating population. Each minute-scale cycle of the dialogue process includes four phases:
\begin{enumerate}
    \item[(i)] An open ended question is sent by the dialogue moderators.
    \item[(ii)] The participating population responds with natural language answers then votes on other participants’ responses.
    \item[(iii)] Results are computed which include the fraction of the participating population predicted to agree with each answer and the confidence of each prediction.
    \item[(iv)] The dialogue moderators review the results and derive learnings which can inform the next dialogue cycle.
\end{enumerate}

\textbf{Model.}
During phase (ii) of the RLSDP we employ two types of voting exercise: agreement and pair choice. In the first, a participant is shown a response and asked if they agree with it. We denote the the event that participant $i$ reports agreement with response $j$ by $a_{ij}$, and disagreement by $d_{ij}$. In the second, the participant is shown two responses and asked which they prefer. We denote the event that participant $i$ reports preference of response $j$ over response $k$ by $c_{ijk}$.  The minute-scale timing of the cycle constrains the total number of exercises per person to be on the order ten. However, the number of participants and thus the number of expected responses is much larger. As a consequence, the first challenge is reminiscent of compressed sensing: given a sparse sampling of exercise data, maximize accuracy in predicting the fraction of participants which agree with each response.

As a baseline, we consider the representative agent model of choice where each response has a utility (independent of the participant) $m_j$, so that $p(a_{ij} | m_j) = \sigma(m_j) = q_j$ \ [13]. We will refer to this as the binomial model of choice. A prominent example is of individual-level choice is probabilistic matrix factorization (PMF) \ [14], popularized by its success in the famous Netflix prize competition. In this context, choice is modeled at the individual level, at the cost of making strict assumptions on the number of factors involved in decision making. This amounts to an upper bound on the rank of the participant-to-item utility matrix. A closely related approach, matrix completion, drops these assumptions and instead uses the low-rank promoting nuclear norm prior. \ [15]

Letting $A = \{i, j \, | \, a_{ij}\}, D=\{i,j \, | \, d_{ij}\}, C = \{i,j,k \, | \, c_{ijk} \}$, and denoting the logistic function by $\sigma$, yields the likelihood
\[
p(A, D, B \, | \, M) = \prod_{i,j \in A} \sigma(m_{ij} + b_i) \prod_{i,j \in D} (1-\sigma(m_{ij} + b_i)) \prod_{i,j,k \in C} \sigma(m_{ij}-m_{ik}).
\]

We now define the nuclear norm $||\cdot||*$. Let $X \in \mathbb R^{d_1 \times d_2}$ be any arbitrary matrix with singular value decomposition $X = U \Sigma V^*$. Then $||X||* := \mathrm{tr}(\Sigma)$ is the sum of the positive singular values $\sigma_1 \geq \sigma_2 \geq \dots \geq \sigma_r > 0$ of $X$. As insig for nuclear norm minimization in low-rank matrix completion, we remark that one may also view $||\cdot||_*$ as the L1 norm of the vector of singular values and thus one may roughly translate the sparsity inducing effects of L1 regularization into rank minimizing effects of the nuclear norm.

We apply a uniform prior over the nuclear norm ball of radius $\tau$. Hence the posterior is
\[
    p(M \, | \, A, D, B) = 
    \frac{1}{Z}
    p(A, D, B \,|\, M) \cdot
    \mathbf{1}_{||M||* < \tau}
\]

\[
    Z := \int_{
            \mathbb R^{n \times m}
        }
        p(A, D, B \,|\, M) \cdot
        \mathbf{1}_{
            ||M||* < \tau}
         \mathrm{d}M.
\]

The second challenge is to estimate the posterior variance needed to compute confidence in predictions within the live timescale. Sufficiently simple models such as the binomial model may admit exact computation of posterior variance. In the case of more realistic models of individual choice, it is necessary to make an approximation. The canonical approaches to approximation of posterior variance and other probabalistic quantities of interest are variational inference \ [16], and Markov chain Monte Carlo methods (MCMC) \ [17]. Recently, stochastic weight averaging (SWA) \ [18] has emerged as a promising local approximant to the posterior. Here, we compare SWA and MCMC methods to obtain confidence estimates in the context of a RLS dialog.
For sampling techniques, it suffices to instead work with the unnormalized posterior $p(A, B \ | \ M)\mathbf{1}{||M||* < \tau}$. The stochastic weight averaging begins from an MLE or MAP estimator, and runs stochastic gradient descent at a high learning rate $\eta = 1$. The iterates are taken as a local surrogate for samples from the posterior. 

\section{Data and Implementation}
Experiments were conducted in a low risk environment. Data was obtained from a sample of 110 participants solicited through the Amazon Mechanical Turk crowdworking platform. The task description was "participate in a live online conversation about your experience as a crowdworker." Participants were instructed to "write a brief, thoughtful response, and evaluate and compare each others' responses." They were asked five questions [A1]. For each question, an average of 136 responses, 1537 agreement exercises, and 1453 pair-choice exercises were collected over the course of 4 minutes.

We implemented the SWA model in PyTorch, and used a Hamiltonian Monte Carlo (HMC) model with the same loss function using Tensorflow and the Tensorflow probability library. As a proxy for repeating the data collection with the same questions and varying duration, the data was split fifty times into training sets including between 5\% and 95\% of the data. We used samples from each model to approximate the posterior standard deviation of the marginal distribution of agreement for each response $\frac{1}{N} \sum_{i} \sigma(m_{ij})$. We analytically computed posterior standard deviation for the binomial model with an uninformative $\mathrm{Beta}(\frac{1}{2}, \frac{1}{2})$ prior for agreement with each response.

\section{Results}

We compute the accuracy of the model on a holdout set of agreement voting exercise as the total data-points per participant used to train the model is varied [figure 1(a)]. An equal number of agreement and choice tasks was determined to be near optimal [A2] and thus used at training time. Accuracy increases linearly initially and the levels off at around 15 data points per participant to about 70-80\% using SWA. We note that the parameters obtained via SWA are significantly better predictors than their HMC counterparts over all ranges, most likely due to their relative proximity to the MLE. 

Confidence is probed by computing the posterior standard deviation of the marginal distribution of population agreement for each response. We compare our model trained using SWA and HMC with a representative agent baseline over a range of data points per participant [figure 1(b)]. Over the entire range explored our model significantly outperforms the baseline for both training methods. At 15 data points per person our model yields a one statndard deviation confidence range of $\pm 1.5 \%$. The average mean absolute error (MAE) between confidence from SWA and HMC is computed for different ranges of data points per person [table 1]. At less than 2 data points per person SWA appears to significantly overestimate confidence. As the number of data points per person increase confidence estimates from SWA and HMC converge, with mean absolute error converging to about $0.2 \%$ at 15 data points per participant. Over this range, SWA runtimes are observed to be less than $1/100^{th}$ as long as those from HMC.

\begin{figure}
\centering
\caption{Training results as a function of data per participant. Ranges shown for validation accuracy and posterior standard deviation. Means denoted as solid line.}
\begin{subfigure}[b]{0.4\textwidth}
\centering
\caption{Model validation}
\begin{tikzpicture}
  \begin{axis}[
  xmin=130, 
  xmax=2400,
  ytick={0.5, 0.6, 0.7, 0.8},
  yticklabels={%
  $50\%$,
  $60\%$,
  $70\%$,
  $80\%$},
  xtick={518.8, 1037.5, 1556.2, 2075.0},
  xticklabels={%
  $5$,
  $10$,
  $15$,
  $20$},
  legend entries = {SWA, HMC},
  legend pos = south east,
  xlabel = data per participant,
  ylabel = validation accuracy
  ]
    \addplot[name path=low, opacity=0, forget plot] table[x=x, y=y_low, mark=none, col sep=comma] {figure1s.csv};
    \addplot[name path=high, opacity=0, forget plot] table[x=x, y=y_high, mark=none, col sep=comma] {figure1s.csv};
    
    \addplot[blue, opacity=.35, forget plot] fill between[of=low and high];
    
    \addplot[blue, thick] table[x=x, y=y, mark=none, col sep=comma] {figure1s.csv};

    \addplot[name path=low, opacity=0, forget plot] table[x=x, y=y_low, mark=none, col sep=comma] {figure1h.csv};
    \addplot[name path=high, opacity=0, forget plot] table[x=x, y=y_high, mark=none, col sep=comma] {figure1h.csv};
    
    \addplot[orange, opacity=.35, forget plot] fill between[of=low and high];
    
    \addplot[orange, thick] table[x=x, y=y, mark=none, col sep=comma] {figure1h.csv};

  \end{axis}
  
\end{tikzpicture}
\end{subfigure}
\hspace{1cm}
\begin{subfigure}[b]{0.4\textwidth}
\centering
\caption{Confidence comparison}
\begin{tikzpicture}
  \begin{axis}[
  xmin=130, 
  xmax=2400,
  ytick={0, 0.05, ..., 0.25},
  yticklabels={%
  $0\%$,
  $5\%$,
  $10\%$,
  $15\%$,
  $20\%$},
  xtick={518.8, 1037.5, 1556.2, 2075.0},
  xticklabels={%
  $5$,
  $10$,
  $15$,
  $20$},
  legend entries = {SWA, HMC, Binomial},
  legend pos = north east,
  xlabel = data per participant,
  ylabel = posterior std. dev
  ]

    \addplot[name path=low, opacity=0, forget plot] table[x=x, y=y_low, mark=none, col sep=comma] {figure2s.csv};
    \addplot[name path=high, opacity=0, forget plot] table[x=x, y=y_high, mark=none, col sep=comma] {figure2s.csv};
    
    \addplot[blue, opacity=.35, forget plot] fill between[of=low and high];
    
    \addplot[blue, thick] table[x=x, y=y, mark=none, col sep=comma] {figure2s.csv};

    \addplot[name path=low, opacity=0, forget plot] table[x=x, y=y_low, mark=none, col sep=comma] {figure2h.csv};
    \addplot[name path=high, opacity=0, forget plot] table[x=x, y=y_high, mark=none, col sep=comma] {figure2h.csv};
    
    \addplot[orange, opacity=.35, forget plot] fill between[of=low and high];
    
    \addplot[orange, thick] table[x=x, y=y, mark=none, col sep=comma] {figure2h.csv};

    \addplot[name path=low, opacity=0, forget plot] table[x=x, y=y_low, mark=none, col sep=comma] {figure2b.csv};
    \addplot[name path=high, opacity=0, forget plot] table[x=x, y=y_high, mark=none, col sep=comma] {figure2b.csv};
    
    \addplot[green, opacity=.35, forget plot] fill between[of=low and high];
    \addplot[green, thick] table[x=x, y=y, mark=none, col sep=comma] {figure2b.csv};
    
    \addplot[green, thick, dashed, forget plot] table[x=x_proj, y=y_proj, mark=none, col sep=comma] {figure2b_proj.csv};

  \end{axis}
  
\end{tikzpicture}
\end{subfigure}

\end{figure}

\begin{table}
\centering
  \caption{Comparison of runtimes and mean absolute error (MAE) between confidence estimates for SWA and HMC across ranges of data per participant (DPP).}
  \vspace*{3mm}
  \begin{tabular}{llll}
    \toprule
    \multicolumn{4}{c}{Mean runtime ($s$)}                   \\
    \cmidrule(r){2-3}
    DPP     & SWA & HMC & MAE \\
    \midrule
    \num{2.5}-\num{5} 	&\num{9.44} 	&\num{356.31}     &\num{9.63e-3} \\
    \num{5}-\num{7.5} 	&\num{12.33} 	&\num{567.74}  	&\num{7.10e-3} \\
    \num{7.5}-\num{10} 	&\num{11.10} 	&\num{833.41} 	&\num{5.95e-3} \\
    \num{10}-\num{12.5} 	&\num{10.45} 	&\num{1186.67} 	&\num{4.77e-3} \\
    \num{12.5}-\num{15} 	&\num{11.28} 	&\num{1216.39} 	&\num{3.18e-3} \\
    \num{15}-\num{17.5} 	&\num{10.53} 	&\num{1407.59} 	&\num{1.90e-3} \\
    \num{17.5}-\num{20} 	&\num{9.82} 	&\num{1809.92} 	&\num{2.03e-3} \\
    \num{20}-\num{22.5} 	&\num{9.71} 	&\num{1881.24} 	&\num{1.96e-3} \\
    \bottomrule
  \end{tabular}
\end{table}
\section{Conclusions, Risks, and Policies}

We conclude that at 15 data points per person our ML model predicts the fraction of a participating population which agrees with a response with reasonably high confidence and that SWA can be used to reliably estimate confidence in seconds. This translates to feasibly executing phases ii and iii of an RLSDP in about 2 minutes.
At this timescale, many cycles of an RLSDP can reliably take place over the course of a one hour dialogue process. Our next step is to pilot an RLSDP in the context of a mediation process, and impose de-risking policies, before deploying it into mediation of active conflicts. If successful deployment of an RLSDP is able to decrease the time to resolve one major conflict by one week, we estimate it could save 230 lives and \$380 million in economic loss [A3].

Any deployment of a realtime large-scale synchronous dialogue process (RLSDP) warrants managing at least three categories of risk. The first risk is inaccurate results due to non-representative data. This can result from bias in questions, disengaged participants, or a participating population which is not representative. The latter can be the result of poor sampling or malicious actors. A policy to manage this first risk may include requirements that (a) dialogue moderators be trained in asking unbiased questions, (b) an appropriate population sampling and participant validation scheme be applied, and (c) randomized human verification of general data quality be conducted. The second risk is inaccurate results due to poor performance of a prediction model. This can result from a faulty machine learning model or programming errors.  A policy to manage this risk may include the requirement for appropriate model verification to take place on production deployments. The third risk is inaccurate conclusions due to misinterpretation of results. This can happen because the results are interpreted in the absence of proper context or confidence in the results are miscalibrated. A policy to manage this risk may include the requirements that (a) proper context be identified and integrated into the interpretation of results and (b) all ML-based results include estimates of confidence.

\newpage

\section*{References}
\small
[1] United Nations; World Bank. 2018. Pathways for Peace : Inclusive Approaches to Preventing Violent Conflict. Washington, DC: World Bank. © World Bank. https://openknowledge.worldbank.org/handle/10986/28337 License: CC BY 3.0 IGO.

[2] Geneva Declaration Secretariat (2008). Global Burden of Armed Violence. {\it Geneva Declaration}, Available from: https://www.refworld.org/docid/494a455d2.html (accessed 6 September 2019)

[3] Iqbal, Z. (2006). Health and Human Security: The Public Health Impact of Violent Conflict. {\it International Studies Quarterly} Vol. 50, No. 3 (Sep., 2006), pp. 631-649

[4] Wagner, Z., Heft-Neal, S., Bhutta, Z. A., Black, R. E., Burke, M., \& Bendavid, E. (2018). Armed Conflict and Child Mortality in Africa: A Geospatial Analysis. {\it The Lancet}. doi:10.1016/s0140-6736(18)31437-5 

[5] Kadir, A., Shenoda, S., Goldhagen, J., Pitterman, S.\ (2018). The Effects of Armed Conflict on Children: Section on International Child Health. {\it Pediatrics Dec 2018}, 142. (6) e20182586; doi:10.1542/peds.2018-2586

[6]  Bircan, C., Brück, T., Vothknecht, M. (2010) Violent conflict
and inequality. {\it IZA Discussion Papers}, No. 4990, Institute for the Study of Labor (IZA), Bonn

[7] Institute for Economics \& Peace. (2018). The Economic Value of Peace 2018: Measuring the Global Economic Impact of Violence and Conflict, Sydney. Available from: http://visionofhumanity.org/ (accessed 6 September 2019)

[8] United Nations, Charter of the United Nations, 24 October 1945, 1 UNTS XVI, available at: https://treaties.un.org/doc/publication/ctc/uncharter.pdf (accessed 06 September 2019)

[9] Strengthening the role of mediation in the peaceful settlement of disputes, conflict prevention and resolution, UN General Assembly, 66th sess, Agenda item 34 (a), UN Doc A/66/811 (25 June 2012), available at: https://peacemaker.un.org/sites/peacemaker.un.org/files/SGReport\_Strenghtening-\\theRoleofMediation\_A66811\_0.pdf (accessed 06 September 2019)

[10] United Nations Department of Political and Peacebuilding Affairs and Centre for Humanitarian Dialogue (2019). Digital Technologies and Mediation in Armed Conflict. 

[11] Rudin, C. (2018). Please Stop Explaining Black Box Models for High-Stakes Decisions. {\it ArXiv}, abs/1811.10154v2

[12] Heinrich, J., Kim, B., Guan, M.Y., Gupta, M. (2018). To Trust Or Not To Trust A Classifier. {\it 32nd Conference on Neural Information Processing Systems (NeurIPS 2018)}, Montréal, Canada

[13] Daniel L. McFadden – Prize Lecture. NobelPrize.org. Nobel Media AB 2019. Fri. 6 Sep 2019. <https://www.nobelprize.org/prizes/economic-sciences/2000/mcfadden/lecture/>

[14] Salakhutdinov, R. \& Mnih., A. (2007). Probabilistic Matrix Factorization. In Proceedings of the 20th International Conference on Neural Information Processing Systems (NIPS'07), J. C. Platt, D. Koller, Y. Singer, and S. T. Roweis (Eds.). Curran Associates Inc., USA, 1257-1264.

[15] Maddox, W., Garipov, T., Izmailov, P., Vetrov, D.P., \& Wilson, A.G. (2019). A Simple Baseline for Bayesian Uncertainty in Deep Learning. {\it ArXiv}, abs/1902.02476.

[16] Candes, E.J. \& Plan, Y. (2010). Matrix Completion With Noise. {\it Proceedings of the IEEE}, 98. 925 - 936. 10.1109/JPROC.2009.2035722. 

[17] Wainwright, M.J., \& Jordan, M.I. (2008). Graphical Models, Exponential Families, and Variational Inference. {\it Foundations and Trends in Machine Learning} Vol. 1: No. 1–2, pp 1-305.

[18] Salakhutdinov, R. \& Mnih. A. (2008). Bayesian probabilistic matrix factorization using Markov chain Monte Carlo. {\it In Proceedings of the 25th international conference on Machine learning}, (ICML '08). ACM, New York, NY, USA, 880-887. DOI=http://dx.doi.org/10.1145/1390156.1390267

[19] Dupuy, K.\ \& Rustad, S.A.\ (2018). Trends in Armed Conflict, 1946–2017. {\it Conflict Trends}, 5. Oslo: PRIO.

\newpage

\section*{Appendix}
\subsection*{A1}
Summary of data collected for each question including number of participants, text responses submitted, agree/disagree votes collected, and number of paired comparison votes collected. Note that participants could submit multiple text responses per question.
{\renewcommand{\arraystretch}{1.3}
\begin{table}[h]
  \label{sample-table}
  \centering
  \begin{tabular}{p{3cm}llll}
    \toprule
    Question & Participants & Responses & Agree/Disagree & Paired comparisons \\
    \midrule

What is your favorite thing to do in your free time?	& 111	& 108	& 2030	& 1951	\\
What motivates you the most in life, and why?	&108	& 154	& 1489	& 1367 \\
What will be the most important political issue in 5 years?	& 95	& 136	& 1345	& 1262 \\
What could Amazon do to improve your experience on Mechanical Turk?	& 101	& 147	& 1283	& 1232 \\

    \bottomrule
  \end{tabular}
\end{table}
}

\subsection*{A2}
Validation accuracy in predicting agreement voting exercises as the mixture of agreement and pair-choice voting exercises was varied and the total voting exercises was held fix. Ranges shown for validation accuracy with mean denoted as solid line.
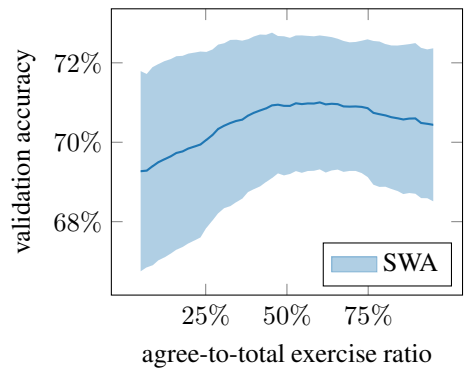
\begin{figure}[h]
\centering
\begin{tikzpicture}
  \begin{axis}[
      ytick = {.68, .7, .72},
      yticklabels={%
      $68\%$,
      $70\%$,
      $72\%$},
      xtick={0.25, 0.5, 0.75},
      xticklabels={%
      $25\%$,
      $50\%$,
      $75\%$},
      legend entries = {SWA},
      legend pos = south east,
      xlabel = agree-to-total exercise ratio,
      ylabel = validation accuracy
    ]

    \addplot[name path=low, opacity=0, forget plot] table[x=x, y=y_low, mark=none, col sep=comma] {figure3s.csv};
    \addplot[name path=high, opacity=0, forget plot] table[x=x, y=y_high, mark=none, col sep=comma] {figure3s.csv};
    
    \addplot[blue, opacity=.35] fill between[of=low and high, forget plot];
    
    \addplot[blue, thick] table[x=x, y=y, mark=none, col sep=comma] {figure3s.csv};

  \end{axis}
  
\end{tikzpicture}
\end{figure}

\subsection*{A3}
Taking 100k people killed in conflict per year [1] plus the average of 5 times more indirect deaths [2] gives a total of 600k deaths per year due to conflict.  We note \$1 trillion in economic loss per year [7] and 49 conflicts per year [19]. This means the average conflict causes about 12k deaths and \$20b in economic loss per year or about 230 deaths and \$380m per week.
\end{document}